\documentstyle[epsfig]{lamuphys}
\makeatletter
\let\chapter\hid@chapter
\makeatother
\newcommand{\be}{\begin{equation}}
\newcommand{\ee}{\end{equation}}
\newcommand{\ba}{\begin{eqnarray}}
\newcommand{\ea}{\end{eqnarray}}
\begin{document}
\pagenumbering{arabic}
\thispagestyle{empty}
\setcounter{page}{0}
\begin{flushright}
LU TP 97/26\\
hep-ph/9710341\\
October 1997
\end{flushright}
\vfill
\begin{center}
{\bf Goldstone Boson Production and Decay\footnote{Invited plenary talk
in the Chiral Dynamics Workshop, Sept. 1-5, 1997, Mainz, Germany}}\\[2cm]
\large
Johan Bijnens\\[0.5cm]
Department of Theoretical Physics 2, University of Lund\\
S\"olvegatan 14A, S22362 Lund, Sweden
\end{center}
\vfill

\begin{abstract}
Various topics in and around Goldstone Boson Production and Decay in
CHPT are discussed, in particular I describe some of the progress
in $p^6$ Chiral Perturbation Theory Calculations, the progress in
calculating hadronic contributions to the muon anomalous magnetic moment,
here comparing the two latest results of the light-by-light in some detail.
I also present some progress in
various $\eta$ and $K$ decays and their relevance for CHPT.
\end{abstract}
\vfill
\newpage
\title{Goldstone Boson Production and Decay}

\author{J.~Bijnens}

\institute{Department of Theoretical Physics 2, University of Lund,\\
S\"olvegatan 14A, S22362 Lund, Sweden}

\maketitle

\begin{abstract}
Various topics in and around Goldstone Boson Production and Decay in
CHPT are discussed, in particular I describe some of the progress
in $p^6$ Chiral Perturbation Theory Calculations, the progress in
calculating hadronic contributions to the muon anomalous magnetic moment,
here comparing the two latest results of the light-by-light in some detail.
I also present some progress in
various $\eta$ and $K$ decays and their relevance for CHPT.
\end{abstract}
\section{Introduction}

Most of the other talks at this conference contained a rather well
defined topic. This talk was left somewhat undefined and I have therefore
taken the liberty of covering topics where there has been a lot of progress
since the last Chiral Dynamics Workshop and which were not covered by any of
the other plenary talks.

Chiral Dynamics and, espescially, Chiral Perturbation Theory (CHPT)
are the main topic in this meeting. It has been introduced by
J\"urg Gasser (\cite{Gasser}) and the variant relevant for the case
of a small quark condensate by Jan Stern (\cite{Stern}). In this talk I
will only cover the standard case. See Stern's talk for references to
the nonstandard case.

There is also a large overlap between this talk and the presentation of the
working group with the same name (\cite{Bijnens}). I will refer to that talk
whenever appropriate. This talk consists of 3 main parts : an overview of
the progress in CHPT at order $p^6$ in the mesonic sector, a discussion
of the relevant chiral dynamics for the hadronic contributions to the muon
anomalous magnetic moment and a few selected $K$ and $\eta$ decays.

In section \ref{2loop} I discuss the presently done full two-loop
calculations. In the two flavour, up and down quark, sector there exist
quite a few calculations. The $\pi\pi$ scattering amplitude has been
discussed both in a plenary talk (\cite{Ecker}) and by several contributions
in one of the working groups (\cite{WG2}). I therefore restrict myself to
the three other calculations: $\gamma\gamma\to\pi^+\pi^-$(\cite{Burgi}),
$\gamma\gamma\to\pi^0\pi^0$(\cite{BGS}) and $\pi\to e\nu\gamma$(\cite{BT}).
In the three flavour case there exists calculations of the vector and
axial-vector two-point functions(\cite{GK1,GK2,GK3}) and of a combination
of vector form factors corresponding to Sirlin's theorem(\cite{Post}).

Sect. \ref{g-2} discusses the light-by-light scattering hadronic contribution
to the muon anomalous magnetic moment. Here there are two recent calculations,
(\cite{HKS3}) and (\cite{BPP2}). I compare the latest numbers of the
various sub-contributions in both calculations and their estimated errors.
The main remaining differences are in the way errors are included and in
the estimate of one contribution where there is a large remaining model
dependence.

The next section discusses a few Kaon decays. Here I will concentrate on
the decays where chiral dynamics plays a large role. This section is
basically a summary of my own and A.~Pich's
talk in the meeting on $K$-Physics in Orsay, June 1996 (\cite{Orsay}).

Section \ref{eta} concentrates on two $\eta$ decays. $\eta\to\pi\pi\pi$
as a test of chiral dynamics and as input for one quark mass ratio,
and $\eta\to\pi^0\gamma\gamma$ as a window on high order CHPT contributions.

The last section summarizes the situation as reviewed in this talk.

\section{Progress in the Mesonic Sector at order $p^6$}
\label{2loop}
\subsection{Two-flavour Calculations}
\subsubsection{$\pi\pi$}
As remarked earlier this has been covered by the plenary talk of G.~Ecker
(\cite{Ecker}) and in more detail by the contributions of Mikko Sainio
and Marc Knecht in the $\pi\pi$ and $\pi N$ working group (\cite{WG2}).
\subsubsection{$\gamma\gamma\to\pi^+\pi^-$}

\begin{figure}
\begin{center}
\epsfig{width=12cm,file=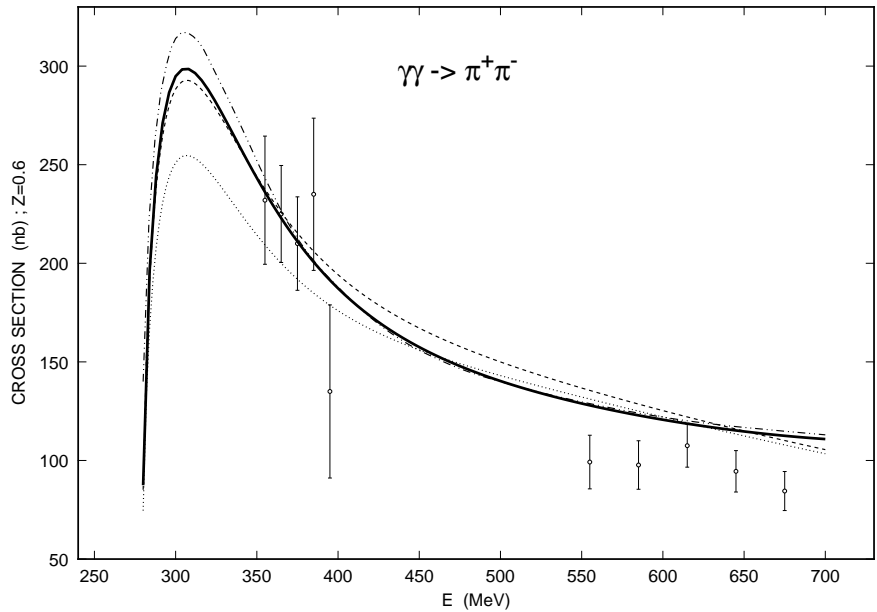}
\end{center}
\caption{The cross section for $\gamma\gamma\to\pi^+\pi^-$.
Figure taken from (B\"urgi 1996).}
\label{fig1}
\end{figure}

The Born term is the same as tree level scattering in Scalar Electro Dynamics
and is known since a long time(\cite{Brodsky}). Early experiments indicated
a large enhancement near threshold over the Born approximation (\cite{PLUTO}).
To order $p^4$ there is one combination of coupling constants that
contributes, $L_{9}+L_{10}$ and there is also a loop contribution(\cite{BC}).
These do provide an enhancement around threshold but as not as large
as (\cite{PLUTO}) indicated. These results were also used to get at the
pion polarizabilities as discussed by (\cite{Holstein}).
The $p^6$ calculations were performed by U.~B\"urgi (\cite{Burgi}).
The number of diagrams is rather large but the final numerical difference
is rather small. In Fig. \ref{fig1} I show the more recent
data (\cite{MarkII})
which do not indicate a large threshold enhancement, the Born, $p^4$ and $p^6$
result. The dotted line is the Born cross-section, the dashed line the
$p^4$ result and the full line the $p^6$ contribution. The
data shown are the Mark II data (\cite{MarkII}). The dispersive estimate
of Donoghue and Holstein is the dashed-double-dotted line (\cite{DH}).
\subsubsection{$\gamma\gamma\to\pi^0\pi^0$}

If we would have used current algebra, we would have gotten a very good
``low energy theorem'' for this process. The $p^2$ contribution
obviously vanishes and there is also no contribution at
order $p^4$, for a modern proof in CHPT see (\cite{BC}). The first
contribution would have come from terms like
$tr\left(\nabla_\mu U F^L_{\alpha\beta}\nabla^\mu U^\dagger F^{R\alpha\beta}
\right)$. If we take the naive order of magnitude for the coefficient
of this type of terms we would have obtained a cross-section of a few
hundredths
of a nanobarn. However, in this case it was obvious that the leading
contribution would come from charged pion rescattering in the final state.
When this process was calculated in CHPT to order $p^4$ this was
also what was found (\cite{BC,DHL}). The cross-section predicted in this
fashion was found to be a few nanobarn. The experimental measurement
afterwards (\cite{CB}) obtained a cross section of this order but disagreed
in shape and was higher near threshold. This disagreement could be
understood in dispersive treatments (\cite{Pennington,DH}).

\begin{figure}
\begin{center}
\epsfig{angle=-90,width=12cm,file=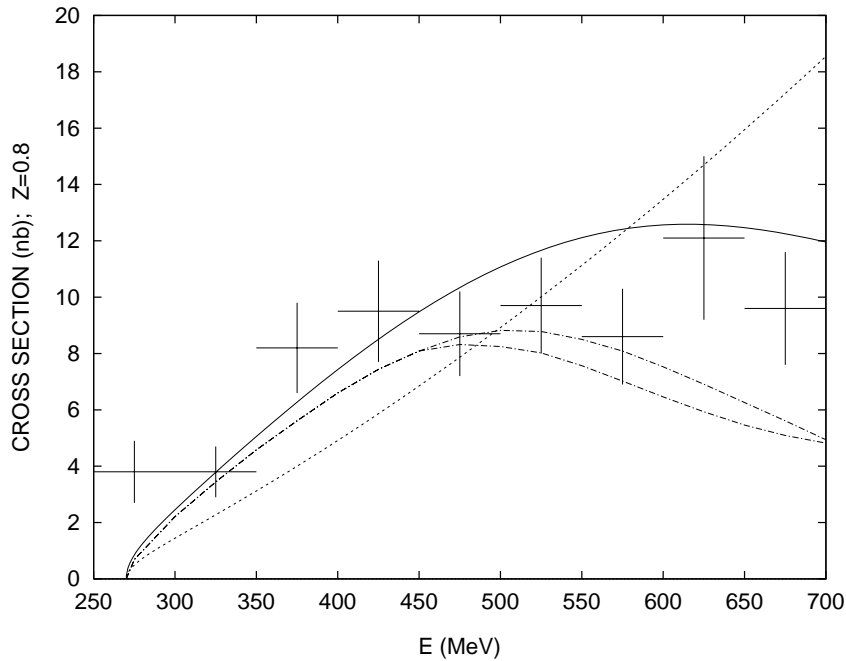}
\end{center}
\caption{The Crystal Ball data, the $p^4$ and the $p^6$ CHPT calculation
as well as the band from dispersive calculations for
$\gamma\gamma\to\pi^0\pi^0$. Figure taken from (Bellucci 1994).}
\label{fig2}
\end{figure}

The calculation at order $p^6$ was the first full two-loop calculation
in CHPT (\cite{BGS}) and showed the same near threshold enhancement.
The result is shown in Fig. \ref{fig2}.
The reason for the large enhancement near threshold was obvious in the
dispersive calculations. At tree level, there is large cancelation between
the $I=0$ and $I=2$ amplitudes. The charged pion cross section has a positive
interference and the neutral pion cross section vanishes. The two isospin
final states have quite different final state interactions which are not too
well described by tree level CHPT. This tree level is the contribution in
the one-loop result while both the $p^6$ result and the dispersive
calculations have a larger $I=0$ $\pi\pi$ final state rescattering
than the tree level result for it. The final state rescattering thus interferes
with the large cancelation present in the neutral pion production, and
while both amplitudes have fairly small corrections, as can be seen
in the charged pion corrections, the sum can have large corrections.

Both the dispersive calculations and the $p^6$ result agree with the Crystal
Ball. The physical effect that creates the bending over towards the higher
center of mass values is the same in both cases as well. It is the
exchange in the $t$-channel of vector mesons. In the CHPT calculation
this comes in via the estimate of the $p^6$ constants while in the
dispersive calculations the vector meson contribution enters via
the so-called left-hand cut.
\subsubsection{Radiative Pion Decay or $\pi^+\to \ell^+\nu\gamma$}

This process serves as the input process for the combination $L_{9}+L_{10}$
used earlier but it is also interesting in its own right. There
are claims that the data cannot be explained by the standard $V-A$ description
of semi-leptonic weak decays (\cite{Chizov}). The same data could
have been explained by an unusually large momentum dependence of one
of the form factors involved in this decay. This decay has three different
contributions, there is the inner Bremsstrahlung-component, which
is by definition given to all orders in CHPT in terms of the pion
decay constant $F_\pi$ and there are two structure dependent form factors.
The vector form factor is given to lowest order by the anomaly and is
known to $p^6$ (\cite{ABBC}). Here the $p^6$ calculation is only
a one-loop calculation. For the pion case there are only small corrections.
The axial-vector form-factor has at $p^4$ only a tree level contribution
(\cite{GL1}), but at two-loop order the loops do contribute (\cite{BT}).

The estimate of the relevant $p^6$ constants is given by axial-vector exchange
and turns out to be very small in the relevant phase space.
Using the standard values of the renormalized couplings at the
$\rho$-mass a sizable correction to the $p^4$ results is found. The uncertainty
due to the uncertainty on the combination $2l_1-l_2$ is smaller than the
uncertainty due to the choice of renormalization scale. The correction is
not negligible despite the fact that the leading correction, the terms
proportional to $L^2=\log^2(m_\pi/\mu)$ vanish in this case.
The size of the various contributions are given in Table \ref{table1}
for 3 different subtraction points.
\begin{table}
\caption[]{Some contributions to the axial form factor in $\pi\to e\nu\gamma$
in units of $10^{-2}\cdot GeV^{-1}$.}
\label{table1}
\begin{center}
\begin{tabular}{lccc}
\hline
$\mu$  & $m_\rho$ & ~0.6~GeV~ & ~0.9~GeV~\\
\hline
${\cal O}(p^4)$ & $-$5.95 & $-$5.95 & $-$5.95\\
$Z_\pi$ and $F\to F_\pi$& $-$0.22 & $-$0.24 & $-$0.21\\
${\cal O}(p^6)$ 1-vertex of ${\cal L}_4$~~ &+1.03 & +0.88 & +1.19\\
${\cal O}(p^6)$ pure two-loop & +0.53 & +0.42 & +0.59\\
\hline
Total & $-$4.62 & $-$4.89 & $-$4.44\\
\hline
\end{tabular}
\end{center}
\end{table}
\subsection{Three flavour results}
The full list of counterterms has been derived for $N_f=3$ by
(\cite{FS}) and for general $N_f$ by Bijnens, Colangelo and Ecker.
\subsubsection{The Vector-Vector two-point function}
\begin{figure}
\begin{center}
\epsfig{angle=90,width=12cm,file=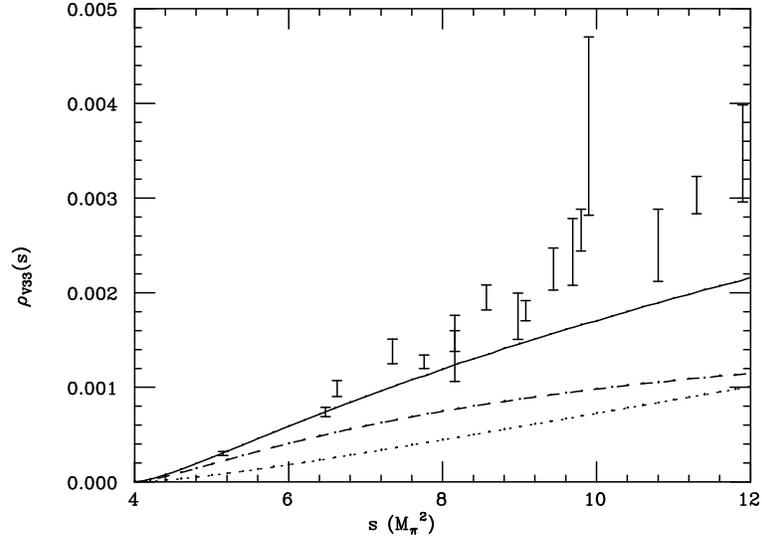}
\end{center}
\caption[]{The vector spectral function at order $p^4$ (dashed),
$p^6$only (dotted) and the sum (full) together with the data from $e^+e^-$.
The buildup of the $\rho$ can be seen here. Figure from
(Golowich and Kambor 1995).}
\label{fig3}
\end{figure}

This has been calculated in (\cite{GK1}) and numerically studied in
more detail in (\cite{GK2}). The quantity to be calculated here is
\be
\Pi^V_{ab}(q^2) = i \int d^4x e^{iq\cdot x}
\langle0|T\left(V_\mu^a (x)V_\nu^b(0)\right)|0\rangle\,.
\ee
The calculation here is simpler since no ``real'' two-loop diagram needs to
be calculated but the complexity of renormalization at two-loops still
hits in its full complexity(\cite{GK1}).
The spectral function from this calculation is shown in Fig. \ref{fig3}.

More important, this calculation can be used in various sum rules.
This leads to predictions for differences of spectral functions
in the up,down and strange sector (here in hyper-charge notation):
\ba
\int_{s_0}^\infty ds \frac{\rho^{33}_V(s)-\rho^{88}_V(s)}{s^{n+1}}
&=& (n=0) \Rightarrow Q=(3.7\pm2.0)~10^{-5}\nonumber\\
&=& (n\ge 1)\Rightarrow L_9^r=(6.6\pm0.3)~10^{-3}\\
\int_{s_0}^\infty ds \frac{\rho^{aa}_V(s)}{s^{n+2}}
&=& (n=0) \Rightarrow P = -(5.6\pm0.6)~10^{-4}\nonumber\\
&=& (n\ge 1) \Rightarrow L_9\mbox{ only}
\ea
The numerical results are taken from (\cite{GK2}). The expressions depend
on 4 constants, $L_9$ and three combinations of $p^6$ constants $P,Q$ and $R$.
The two that can be determined via the sum rules agree well with the resonance
estimate of the same quantities thus providing evidence that this method
for estimating the constants also works at order $p^6$.
They have recently calculated also the $AA$ two-point function
and a similar numerical analysis is under way(\cite{GK3}).
This is discussed
in (\cite{Bijnens}). Other relevant references are the calculation of
(\cite{Holdom}) for a two-loop vector two-point function without the
renormalization aspects and the calculation by Maltman of the
isospin breaking $\langle T(V^3 V^8)\rangle$ vector two-point function
(\cite{Maltman}).
\subsubsection{Sirlin's Theorem}
In (\cite{Sirlin}) it was proven that the combination
\be
\Delta(t)=\underbrace{\frac{1}{2}F^{\pi^+}(t)+\frac{1}{2}F^{K^+}(t)
+F^{K^0}(t)}_{\mbox{Electromagnetic\\Form Factors}}
-\underbrace{f_+^{K\pi}(t)}_{\mbox{weak $f_+$ form-factor}}
\ee
only starts at order $(m_s-\hat{m})^2$. An immediate consequence of this is
that at $t=0$ dependence on $p^6$ parameters exists, e.g. via
terms of the type $\langle u_\mu u^\mu\chi_+^2\rangle$. But by
powercounting, the $t$-dependence of these form-factors is at most
$t (m_s-\hat{m})$, and the charge radius of the above combination thus
has no contribution from terms in $p^6$-Lagrangian. Caution must be taken here,
the combination $\Delta(t)$ has large cancelations in it and we can thus
expect large higher order corrections. That CHPT is well behaved for
this quantity can be seen when comparing the size of the $p^6$ correction to
the charge radius of $\Delta(t)$ with the individual charge radii
of the combination.

The result is (\cite{Post})
\ba
\langle r^2\rangle_{\mbox{Sirlin}}& =&
\left(0.006 (p^4) + 0.017(3)(\mbox{reducible})-0.002(\mbox{irr.})\right)~fm^2
\nonumber\\
&=& (0.021\pm0.003)~fm^2\,.
\ea
This should be compared with the experimental results
\ba
\langle r^2\rangle_{\mbox{Sirlin}}\hskip-2cm\nonumber\\
&=&\left(
\frac{1}{2}\left[0.439(8)(\pi^+)+0.34(5)(K^+)\right]
-0.054(26)(K^0)-0.36(2)(K\pi)\right)fm^2
\nonumber\\& =&-(0.025\pm.041)~fm^2\,.
\ea
The size of the $p^6$ correction is less than $10\%$ of the largest terms
so it is a nicely converging result. The present experimental precision
is too low to significantly test this calculation.
\section{Muon Anomalous Magnetic Moment}
\label{g-2}
There is a new experiment on the muon anomalous magnetic moment,
$a_\mu=(g-2)/2$, planned at BNL (\cite{BNL}). They aim at a precision in
$a_\mu$ of $4\cdot10^{-10}$, to be compared with the present precision
of $84\cdot10^{-10}$ from the CERN experiment. The main aim
is to unambiguously detect the weak gauge boson loops and put constraints
on possible other contributions.

We therefore need to determine the contributions from the strong interaction
with great precision. There are three hadronic contributions relevant
at the present level of precision, the hadronic vacuum polarization,
the higher order vacuum polarization and the light-by-light contribution.
The first two depend on the same integral over the hadronic vector spectral
function which can be measured in tau decays and in electron-positron
collisions. The latest determination is in (\cite{Alemany}) and is
also discussed in some detail in (\cite{Bijnens}). Here the main
need is for more precise experiments in the rho mass region in order to
bring the error down to the precision of the BNL experiment.
Theoretical estimates of this quantity are accurate to about 25\%
(\cite{Rafael,Pallante}).

The light-by-light contribution is more of a problem, it cannot be related
to experiments in a simple way and has therefore to be evaluated in a
theoretical framework. The relevant quantity is an hadronic four-point function
so lattice QCD determinations are probably some time into the future.
This quantity has been calculated recently by two groups with the following
recent history: (all in units of $10^{-10}$)
\ba
&&-3.6\pm1.6 \mbox{(\cite{HKS1})}; -5.2\pm1.8 \mbox{(\cite{HKS2}) and }
\nonumber\\&&
-7.9\pm1.5 \mbox{(\cite{HKS3})}
\label{HKS}\\&&
-11\pm5 \mbox{(\cite{BPP1}) and }-9.2\pm3.2 \mbox{(\cite{BPP2}).}
\label{BPP}
\ea
The two results are in fact in quite good agreement with each other
on the total value and on the error but they differ
in the error combining.
The reasons for the change in the numbers are for (\ref{HKS}):
first a change in the model coupling pseudoscalar mesons (P) to
two photons, and for the second change the inclusion of the $\eta'$ and
a small change with the $P\gamma\gamma$ coupling because
of the measured CLEO $P\gamma\gamma^*$ form factor.
For (\ref{BPP}) the change was a change in the $P\gamma\gamma$ coupling
to agree better with the preliminary CLEO data
(following a suggestion of Kinoshita).

The three different type of contributions to the light-by-light diagram,
different in chiral and $N_c$
counting are (in units of $10^{-10}$): first (\ref{HKS}), second (\ref{BPP})
\begin{tabbing}
$\pi^0$, $\eta$ and $\eta'$ exchange
\hskip1cm \= $-8.3\pm0.65$\hskip1cm \= $-8.5\pm1.3$\hskip1cm \= Good\\[0.5cm]
axial+scalar exchange \> $-0.17$ \> $-0.93\pm0.03$ \>\\
quark loop \> $1.0\pm1.1$ \> $2.1\pm0.3$ \>
Good\\*
The ENJL model used for (\ref{BPP})
here tends to mix these two contributions,\\therefore only the sum
can be compared between (\ref{HKS}) and (\ref{BPP}).\\[0.5cm]
charged pion and Kaon loop \> $-0.45\pm0.81$ \> $-1.9\pm1.3$ \>
Main\\*
Model used for loop\> HGS \> naive VMD \>uncertainty\\[0.5cm]
Errors added linearly \> $\pm2.6$ \> $\pm3.2$ \>\\*
Errors added quadratically \> $\pm1.5$ \> $\pm1.9$\\
\end{tabbing}
The pseudoscalar exchange contribution we agree on extremely well.
The error in (\cite{BPP2}) was chosen larger because we only have tested
the models with one photon off-shell, while both photons off-shell
contribute also significantly. For the 2nd contribution the error estimate went
the other way, in (\cite{HKS2}), there is the freedom of the quark mass,
in (\cite{BPP2}) a good matching between long and short distance was observed
and hence a smaller error chosen.
The main difference is really in the last contribution where two different
but equivalently good chiral models were chosen for the relevant
$\gamma^*\gamma^*\pi\pi$ coupling. Both models are chirally invariant and
satisfy the correct chiral identities when the off-shellness is extrapolated
to the rho-meson pole. In my opinion we should therefore choose the error
such that it includes both models.

In conclusion, in order to improve the present situation we need data
on the couplings of one and two pseudoscalar mesons to two photons
with {\em both} photons off-shell.
\section{Kaon decays}
\label{kaon}
This section can be found more extensively in the talks by J.~Bijnens
and A.~Pich in (\cite{Orsay}). $CP$-violation and $\varepsilon'/\varepsilon$
are also covered in great detail in those proceedings and I will therefore
not treat them here. Extensive treatments can also be found
in (\cite{DAPHNE}).
\subsection{Semi-leptonic Kaon Decays}
\subsubsection{$K_{l3}$ : $K\to\pi e\nu$, $K\to\pi\mu\nu$}
the main problem here is that we need improvement in the experimental
situation on the slope of the scalar form factor. It should also be
remembered that these decays are our main source of knowledge of
the Cabibbo angle, or of $V_{us}$, and thus deserve very
accurate measurements.

\subsubsection{$K_{l2\gamma}$}
this decay is similar to the pion radiative decay discussed above and has
similar characteristics. The vector form factor is a test of the anomaly
and an accurate measurement of this would be an independent measurement
of quark mass effects in anomalous amplitudes. At present the only place
where this occurs is in $\eta$ decays and there the question
is entangled with $\eta\eta'$ mixing. The vector Form factor is known
to $p^6$(\cite{ABBC}) assuming very small direct quark mass effects.
The axial vector form factor depends on $L_{9}+L_{10}$ and is thus
predicted from the pion decay. Given the corrections seen there at order $p^6$
the prediction for this form-factor has an expected error of about 30\%.

\subsubsection{$K_{l2ll}$} Here again everything is known to order $p^4$,
there are large enhancement for the modes involving electrons
over the inner Bremsstrahlung amplitude and there is good agreement
with experiment but the experimental errors are fairly large.

\subsubsection{$K_{l4}$} This decay has been discussed in the framework
of obtaining new accurate values of the $\pi\pi$ phase shifts. It should not
be forgotten that the absolute values of the 4 form-factors are themselves
also of interest. They provide additional input for the parameters of CHPT
while at the same time providing tests of
CHPT (\cite{BCG}). By measuring all the channels
one can also test the isospin assumptions underlying the present theory
calculations.

\subsubsection{$K_{l3\gamma}$} The $p^4$ correction is rather small
due to a cancelation between the ``counterterm'' and the loop contributions.
This cancelation is in fact necessary to obtain agreement with the
measurement of $(3.61\pm0.014\pm0.021)\cdot 10^{-3}$
(\cite{Leber}). The tree level,
$p^4$ prediction is 3.6, 3.8 in the same units(\cite{BEG}).
\subsection{$K\pi\pi$, $K\to\pi\pi\pi$}
This has been discussed extensively by Maiani and Paver in (\cite{DAPHNE}).
As was realized in (\cite{DHKW}) the $p^4$ calculations leave in fact
a series of relations between various slope parameters in $K\to3\pi$
when the $K\to2\pi$ and $3\pi$ rates are used as input.
These relations provide stringent tests of Chiral Perturbation Theory
in this sector and need to be tested so our predictions for CP violating
effects can be refined.
At present the agreement is satisfactory but especially in the $\Delta I=3/2$
sector the experimental precision is rather poor.
\subsection{Rare $K$ decays}
This area has been the scene of some of the major successes
of CHPT, but there are also some problem cases.
\subsubsection{$K_S\to\gamma\gamma$}
This is a parameter-free CHPT prediction at order $p^4$(\cite{Kgg}).
The experimental measurement of a Branching Ratio of $(2.4\pm0.9)\times
10^{-6}$ (\cite{Kggexp}) agrees extremely well with the prediction of $2.0
\times10^{-6}$.

\subsubsection{$K_L\to\gamma\gamma$}
This process proceeds through $K_L\to\pi^0,\eta\to\gamma$ and at order
$p^4$ there is an exact cancelation between these two contributions.
As a consequence the predictions is very sensitive to higher order effects
and this decay is not really under theoretical control yet.

\subsubsection{$K_L\to\pi^0\gamma\gamma$}
\begin{figure}
\begin{center}
\epsfig{file=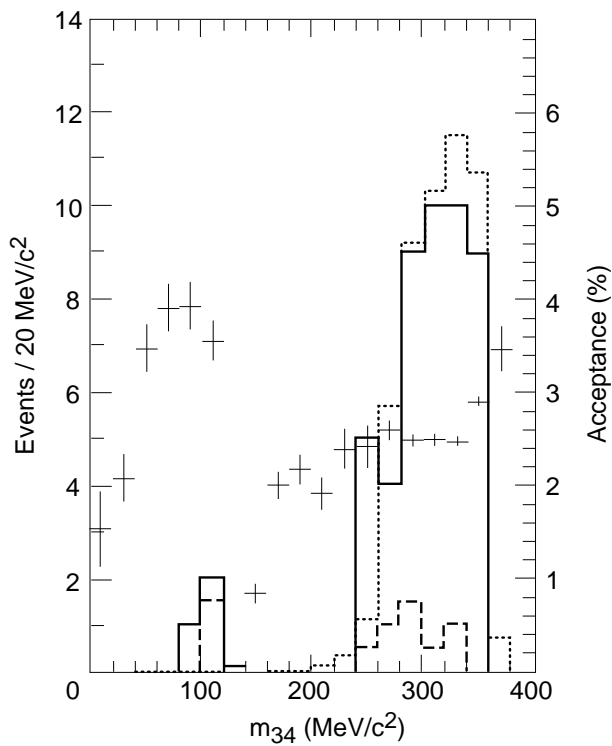,width=8cm,height=10cm}
\end{center}
\caption[]{Measured $2\gamma$-invariant mass
distribution for $K_L\to\pi^0\gamma\gamma$
(solid line). The dotted line simulates the $p^4$ CHPT prediction, normalized,
and the dashed line is the estimated background.
The crosses are the experimental acceptance, figure from  (Barr 1992).}
\label{fig4}
\end{figure}

This decay is also a parameter-free prediction at order $p^4$
(\cite{EPR,CD}). he predicted rate at $p^4$ is a branching
ratio of $0.67\times10^{-6}$ to be compared with
the measured one of $1.70\times10^{-6}$ (\cite{Kpgg}). But the phase space
distribution is clustered at high $\gamma\gamma$ masses contrary to
Vector meson Dominance model predictions and agrees well with the
CHPT prediction as shown in Fig. \ref{fig4}.

There are two $p^6$ effects expected, unitarity corrections and Vector meson
Exchange contributions. The former make the distribution steeper and the latter
flatter and they both increase the rate. It is possible to get
agreement with both the rate and the spectrum with reasonable estimates
of these contributions, see the contribution
by A~Pich in (\cite{Orsay}) and references therein.
\section{$\eta$-decays}
\label{eta}
\subsection{$\eta\to3\pi$}
There are three questions here in the theory :
\begin{enumerate}
\item total rates
\item the ratio $\displaystyle r=\frac{\Gamma(\eta\to\pi^0\pi^0\pi^0)}
{\Gamma(\eta\to\pi^+\pi^-\pi^0)}$
\item the Dalitzplot distributions
\end{enumerate}
\subsubsection{The Decay Rate}
The order $p^2$ contribution was calculated a long time ago by Cronin
and is about $66~eV$. This should be compared with the Particle Data Group
width of $280\pm28~eV$. The $p^4$ corrections were calculated (\cite{GL2})
and were large, leading to about $167\pm50~eV$. There are two reasons for
this large correction: $\eta\eta'$ mixing and final state rescattering.
The former should be adequately dealt with at the $10\%$ level but
the final state corrections could be
large. These have been evaluated independently by two groups
using dispersive methods
(\cite{eta3pi1,eta3pi2}), earlier references can be found in these
papers. The $p^4$ calculation is used to determine the subtraction constants.
The Dalitz plot parameters are used as constraints on the calculation.
This leads to a value of $209\pm20~eV$ for the decay width.
So, with a slightly large value of $(m_d-m_u)/(m_s-\hat{m})$ we reproduce
nicely the observed decay rate. This slightly larger value of that quark
mass ration was in fact expected from calculated large deviations from
Dashen's theorem (\cite{DHW,Bijnens (1993)}) and
a large number of more recent references. One can now turn in fact the
argument around and the decay rate of $\eta\to3\pi$ has become the most
accurate source of information on that quark mass ratio.
Electromagnetic corrections to the decay rate have since been found
to be small as expected from current algebra arguments (\cite{Baur 1996})
and (\cite{Gosdzinsky 1996}).
\subsubsection{$r$}
The lowest order prediction for the ratio $r$ is $1.53$, the $p^4$ prediction
$1.43$ and the dispersive calculations lead to $1.41\pm0.03$.
The Particle Data Group quotes $1.36\pm0.05$. More precise measurements
of this quantity as an important check on the dispersive calculations
are therefore welcome.
\subsubsection{Dalitzplot}
The Dalitzplot is parametrized as $1+a y + b y^2 +c x^2$ for the charged decay
and as $1+ g (x^2+y^2)$ for the neutral decay. The next-to-leading order
prediction (\cite{GL2}) and the dispersive results together with the
available experimental results are given in Table \ref{table2}. The last
line are new results to be published but cited in (\cite{Amsler}).
As is obvious from the table the agreement is reasonable but an increase in
precision is definitely welcome, given that these numbers are important for
determining the quark mass ratio mentioned above.
\begin{table}
\caption[]{Theoretical and Experimental results for the Dalitzplot parameters.
The first two lines are the theoretical results from the $p^4$ and
the dispersive calculations. The others the experimental results.}
\label{table2}
\begin{center}
\begin{tabular}{c@{~~}c@{~~}c@{~~}c@{~~}c}
\hline
 & $a$ & $b$ & $c$ & $g$ \\
\hline
$p^4$ & $-1.33$ & $0.42$ & $0.08$ & 0 \\
Dispersive & $-1.16$ & $0.26\pm0.01$ & $0.10\pm0.01$ & $-0.014\pm0.014$\\
Gormley & $-1.17\pm0.02$ & $0.21\pm0.03$ & $0.06\pm0.05$ & \\
Layter & $-1.08\pm.014$ & &$0.046\pm0.031$ & \\
Amsler 1995 & $-0.94\pm0.15$ & $0.11\pm0.27$ & & \\
Alde & & & & $-0.044\pm0.046$\\
Amsler 1997 & $-1.19\pm0.07$ & $0.19\pm0.11$ & & $-0.104\pm0.040$\\
\hline
\end{tabular}
\end{center}
\end{table}
\subsection{$\eta\to\pi^0\gamma\gamma$}
This decay provides a window on rather high order CHPT effects.
The loop effects at $p^4$ and $p^6$ are suppressed by either heavy intermediate
states or $G-parity$. The first loops where this suppression is not present
are those with double Wess-Zumino vertices at order $p^8$(\cite{Ametller}).
Those are in fact of the same size as the $p^4$ loops.

The main contribution to the decay starts at order $p^6$ as estimated
from Vector Meson Exchange or the ENJL model. The results are given in
table \ref{table3} for various contributions. The different possibilities
are distinguishable in the the $E_\gamma$ spectrum. The present experimental
value for the width is $0.85\pm0.18~eV$ and the theoretical situation has
despite some theoretical effort not really changed since (\cite{Ametller}).
\begin{table}
\caption{Various contributions to $\Gamma(\eta\to\pi^0\gamma\gamma)$ in $eV$.}
\label{table3}
\begin{center}
\begin{tabular}{lc}
\hline
Contribution & $\Gamma$ ($eV$)\\
\hline
$p^4$ & 0.0039\\
$p^6$ VMD & 0.18\\
$p^6$ ENJL & 0.18\\
$p^6$ VMD + scalar +tensor&0.08--0.33\\
$p^4$+$p^6$+$p^8$(WZ-loops)&0.20--0.45\\
VMD all orders & 0.32\\
\hline
\end{tabular}
\end{center}
\end{table}
Adding all contributions in the table leads to
$\Gamma(\eta\to\pi^0\gamma\gamma)=$0.45--0.50~eV with a large uncertainty.
In reasonable but not very good agreement with the experimental value.
A remeasurement of the decay rate and a measurement of the decay distributions
is certainly desirable.
Understanding of this decay is also needed to determine
the rates for $\eta\to\pi^0\ell^+\ell^-$ decays and also plays a role
in $K_S\to\pi^0\gamma\gamma$.
\section{Conclusions}
There is nice progress at the $p^6$ front in various processes.

The light-by-light contribution to the muon anomalous magnetic moment
is understood to the precision needed for the future BNL experiment
but lowering the error by a factor of 2 would be desirable. The latter
requires experimental input on processes with two photons off-shell,
both $\gamma^*\gamma^*\to\pi^0,\eta,\eta'$ and $\gamma^*\gamma^*\to\pi^+\pi^-$.

In Kaon and $\eta$ decays there is a lot of experimental and theoretical
progress. In this talk I have only covered a small part of the possible
decays and CHPT tests in this area.
%

\end{document}